\documentclass[twocolumn,showpacs,preprintnumbers,amsmath,amssymb,prb]{revtex4}
\usepackage{graphicx}
\usepackage{amsmath, amsthm, amssymb}
\usepackage{latexsym}
\usepackage{amssymb}
\usepackage{bm}
\usepackage{dcolumn}
\usepackage{epstopdf}
\usepackage{color}

\newcommand{\et}{{\it et\  al.}}

\begin{document}

\title{Magnetic field dependence of spin-lattice relaxation in the s$\pm$ state of Ba$_{0.67}$K$_{0.33}$Fe$_{2}$As$_{2}$}

\author{Sangwon Oh$^{1}$, A. M. Mounce$^{1}$, W. P. Halperin$^{1}$, C. L. Zhang$^{2}$, Pengcheng Dai$^{2}$, A. P. Reyes$^{3}$, P. L. Kuhns$^{3}$}

\affiliation{$^1$Department of Physics and Astronomy, Northwestern University, Evanston, Illinois 60208, USA \\
$^2$Department of Physics and Astronomy, The University of Tennessee, Knoxville,
Tennessee 37996, USA\\
$^3$National High Magnetic Field Laboratory, Tallahassee, Florida 32310, USA}
\date{Version \today}

\begin{abstract}

The spatially averaged density of states, $\langle$$N(0)$$\rangle$, of an unconventional $d$-wave superconductor is magnetic field dependent, proportional to $H^{1/2}$, owing to the Doppler shift of quasiparticle excitations in a background of vortex supercurrents.~\cite{vol88,vol93}  This phenomenon, called the Volovik effect, has been predicted to exist for a sign changing $s\pm$ state,~\cite{ban10} although it is absent in a single band $s$-wave superconductor.    Consequently, we expect there to be Doppler contributions to the NMR spin-lattice relaxation rate,  $1/T_1  \propto$$\langle$$N(0)^2$$\rangle$, for an $s\pm$ state which will depend on magnetic field.  We have measured the $^{75}$As $1/T_1$ in a high-quality, single crystal of Ba$_{0.67}$K$_{0.33}$Fe$_{2}$As$_{2}$ over a wide range of field up to 28 T.  Our spatially resolved measurements show that indeed there are Doppler contributions to $1/T_1$ which increase closer to the vortex core, with a spatial average proportional to $H^2$,  inconsistent with recent theory.~\cite{ban11}

\end{abstract} 

\pacs{ }

\maketitle

The spin-lattice relaxation rate, $1/T_1$, in the superconducting state can provide valuable information about  gap structure and about the effects of vortices on the quasiparticle density of states.   
There have been a number of reports on the temperature dependence of $1/T_1$ at low magnetic fields for various pnictide superconductors,~\cite{kam08,nak10,nin10,fuk09,yas09,li11,gra08}  which have been interpreted in terms of order parameter structure.  However, impurities of unknown origin and concentration can play an important role in the analysis. For this reason, conclusions about the superconducting state based on temperature dependences can be ambiguous.

An alternative approach to explore unconventional character of the order parameter  is the magnetic field dependence of the density of states, which can be specific to a particular order parameter symmetry, easily probed through specific heat or thermal conductivity measurements.~\cite{yip92}  The unconventional structures of the $s\pm$ and  $d$-wave  states each lead to characteristic magnetic field dependences of the spatially averaged density of states, attributable to the Doppler shift of quasiparticle excitations, known as the Volovik effect.~\cite{vol88,vol93}  For the $d$-wave case $c$-axis line nodes in the gap give rise to  a field dependence, $\propto H^{1/2}$ and a non-linear Meissner effect.~\cite{yip92}  In contrast,  according to Bang,~\cite{ban10} in the case of $s\pm$ symmetry for a multiband superconductor, the spatially averaged density of states at the Fermi energy is proportional to the magnetic field.  Since $1/T_1$ is proportional to the square of the local density of states, one might think that the predicted Volovik effect should be  $\propto H^2$. However, the Volovik effect pertains to the {\it spatial average} over the vortex unit cell which decreases in area inversely proportional to the magnetic field.  According to the theory, there is a region of normal state excitations surrounding the vortex core of radius $\xi(\Delta_2/\Delta_1)$, where $\xi$ is the core radius $(\sim 30\,\, \AA)$, equal to the coherence length,  and  $\Delta_2/\Delta_1$ is the ratio of large to small gaps, leading to the prediction, $1/T_1T \propto H$ at low magnetic fields, $H \leq H_{c2} (\Delta_1/\Delta_2)^2 \sim 3$ T; otherwise it should be constant.~\cite{ban11}    In this Letter, we report  $^{75}$As NMR measurements in single crystals of Ba$_{0.67}$K$_{0.33}$Fe$_{2}$As$_{2}$ covering a wide range of magnetic fields.   Our results show that indeed there is a Doppler contribution to the spatially averaged spin-lattice relaxation but that in the low temperature limit,  $1/T_1T \propto H^2$ over the whole  range of magnetic field, inconsistent with prediction.~\cite{ban11}

We performed our $^{75}$As NMR measurements at Northwestern University and the National High Magnetic Field Laboratory,  from 4 K to room temperature with external magnetic field from 6.4 to 28 T. The fields were  parallel to the $c$-axis of the single crystals, Ba$_{0.67}$K$_{0.33}$Fe$_{2}$As$_{2}$ (BaK122) that had a zero-field $T_c = 38$ K and were grown at the University of Tennessee by the self-flux method.~\cite{zha11} To increase signal intensity in the superconducting state, the crystals were cleaved to dimensions of 3$\times$3$\times$0.1 mm$^{3}$ and total mass of 17 mg. Typically, spin echo sequences ($\pi/2$ - $\pi$) were used to obtain the spectrum, Knight shift, and $1/T_1$ for the central transition (-1/2 $\Leftrightarrow$ 1/2) with a $\pi$-pulse $\approx$ 7 $\mu$sec. The spin-lattice relaxation  was measured with the full recovery method (28 to 300 K) and  progressive saturation techniques,~\cite{mit01a} (4 to 26 K) the latter being more accurate for very long relaxation times at low temperatures. The average rate was measured with the $\pi$-pulse centered on the spectrum.  Frequency-resolved spin-lattice relaxation was also measured by dividing the spectrum into many small frequency windows and the relaxation was determined separately in each window. Knight shift measurements were performed with a frequency sweep method.
 
\begin{figure}[!ht]
%%%%%%%%%%%%%%%%%   F I G U R E  1   %%%%%%%%%%%%%%%%%%
\centerline{\includegraphics[width=0.4\textwidth]{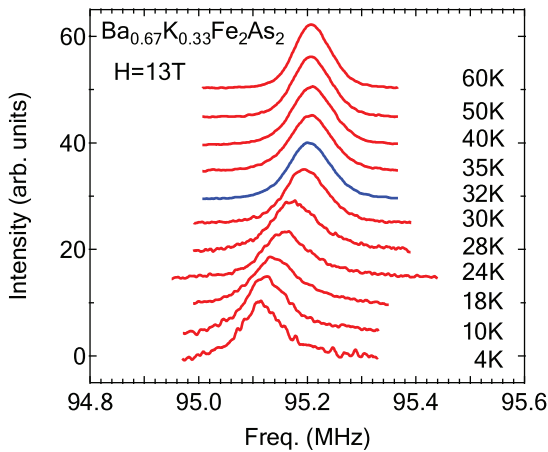}}
%%%%%%%%%%%%%%%%%%%%%%%%%%%%%%%%%%%%%%%%%%%%%
\caption{$^{75}$As NMR spectra of Ba$_{0.67}$K$_{0.33}$Fe$_{2}$As$_{2}$ measured by a frequency sweep technique in 13 T with $H|| c$-axis of the crystals. Below $T_c = 32$ K (blue trace) the spectra shift noticeably to lower frequency with decreasing temperature. The linewidths of the spectra decrease in the superconducting state as reported in some other compounds,~\cite{che07,oh11} where this was attributed to reduction in the local field distribution from impurities.
}
\label{fig1}
\end{figure}

\begin{figure}[!ht]
%%%%%%%%%%%%%%%%%   F I G U R E  2   %%%%%%%%%%%%%%%%%%
\centerline{\includegraphics[width=0.4\textwidth]{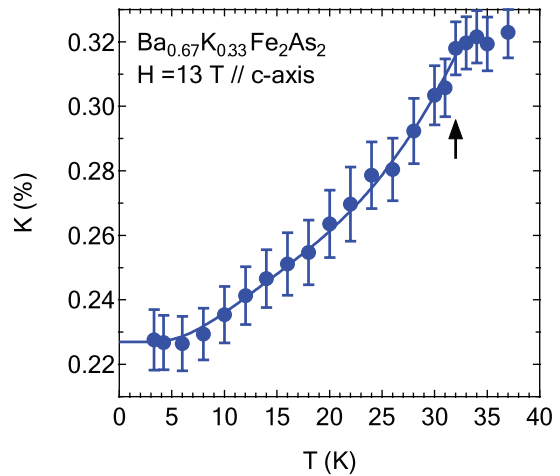}}
%%%%%%%%%%%%%%%%%%%%%%%%%%%%%%%%%%%%%%%%%%%%%
\caption {The total Knight shift K(T) is shown for $H=13$ T where the temperature dependence is associated with  the spin part, $K_s$, that decreases below $T_c$, consistent with spin-singlet superconductivity. The black arrow indicates $T_c = 32$ K.  The data can be fit phenomenologically assuming that the low temperature spin shift is proportional to an average density of states, Eq.~\ref{eq1}, represented by the solid curve, with an orbital shift, $K_{orb}=0.21\%$.}

\label{fig2}
\end{figure}

Early experiments on optimally, hole-doped, single crystals of  Ba$_{0.6}$K$_{0.4}$Fe$_{2}$As$_{2}$~\cite{muk09} grown with tin flux did not detect any signal below 20 K due to linewidth broadening from paramagnetic impurities on the As sites at a level of $\approx 1 \%$. However, there have been substantial improvements in lowering the impurity concentration using the self-flux method.~\cite{li11,zha11} The frequency-swept $^{75}$As NMR spectra of our crystals in 13 T with $H||c$-axis,  are shown in Fig. \ref{fig1}.  The $T_c$ in $H=13$ T is 32 K, and a shift of the spectra can be easily seen. This decrease of the Knight shift indicates spin-singlet pairing in the superconducting state. On cooling, the linewidth slowly broadens from 60 kHz at $T=300$ K to 70 kHz at $T_c$. Below $T_c$ the linewidth increases up to 110 kHz near $20$ K, and then it decreases  to 80 kHz at 4 K, and is  independent of magnetic field from 6.4 to 16.5 T, to within 10\%. The weak dependence of the linewidth on magnetic field and temperature in the normal state indicates that few magnetic impurities are present, comparable to the cleanest cuprate crystals such as Bi$_2$SrCa$_2$Cu$_2$O$_{8+\delta}$ (Bi2212).~\cite{che07}  This point is also consistent with the similar results we find from our comparison of  the zero field extrapolations of $1/T_1$ with those of clean Bi2212 crystals which we discuss later.   
The Knight shift, $K = K_s + K_{orb}$, was determined from the first moment  of the NMR spectrum where $K_s$ and $K_{orb}$ are the spin and orbital parts of the shift, respectively. The orbital part is temperature and field independent, consequently the temperature dependence of the shift in Fig.~\ref{fig2} can be associated with $K_s$, decreasing below $T_c$ on cooling.  The solid curve in the figure is the  temperature dependence of $K_s$ that we describe with a phenomenological model for the density of states, Eq.~\ref{eq1}, based on the parameters obtained from $1/T_1$ measurements.

The behavior  of the spin-lattice relaxation in the superconducting state is the main focus of our present work where we measure the temperature and magnetic field dependence for $H= 6.4, 10.8, 14, 16.5, 27$ and $28$ T, parallel to the $c$-axis of the crystals. The rates were measured with the spectrometer frequency set at the peaks of the spectra.  
A coherence peak below $T_c$ was not observed, and the suppression of $T_c$ by the magnetic field was minimal, from $T=32$ to 30 K when the external field was increased from 6.4 to 27 T.  In low magnetic fields, {\it i.e.} 6.4  and 10.8 T, the temperature dependence of $1/T_1$ could   be approximately described as  $T^3$ at intermediate temperature,  as has often been reported elsewhere.~\cite{gra08,fuk09} But  in higher fields, 14, 16.5, 27 T, below $T=10$ K, we find $1/T_1 \propto T$ , indicating a constant average density of states at zero energy , $\langle$$N(0)$$\rangle$. Recently Li $\et$~\cite{li11} observed an exponential temperature dependence of the rate in a magnetic field of $H=7.5$ T consistent with the presence of a full gap.  A comparison of our data with that of Li $\et$ shows that they are identical except at our lowest temperature, $T=4$ K, where our higher value of $1/T_1$ might be understood as the effect of residual  impurities in our crystal obscuring exponential behavior.  Increasing the magnetic field we find that the spin-lattice relaxation at 4 K increases systematically indicating the existence of a field dependent  density of states at the Fermi surface.  This observation is a  characteristic signature of a Volovik effect.
\begin{figure}[!ht]
%%%%%%%%%%%%%%%%%   F I G U R E  3   %%%%%%%%%%%%%%%%%%
\centerline{\includegraphics[width=0.4\textwidth]{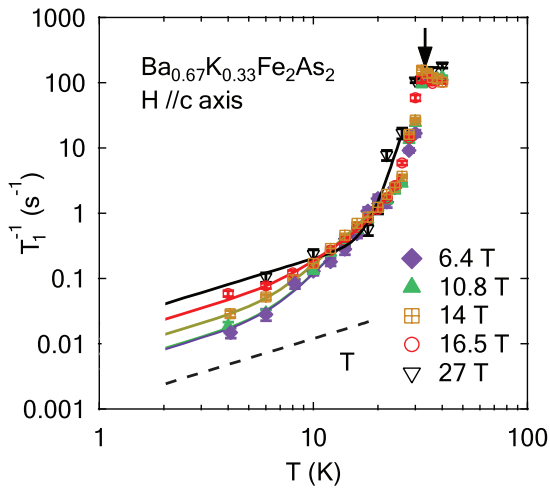}}
%%%%%%%%%%%%%%%%%%%%%%%%%%%%%%%%%%%%%%%%%%%%%
\caption{The spin lattice relaxation rate, $1/T_1$, of Ba$_{0.67}$K$_{0.33}$Fe$_{2}$As$_{2}$ in magnetic fields, $H=6.4, 10.8, 14, 16.5$, and 27 T. The rates were measured at the peak position in the spectrum. The data is consistent with our two-gap model (solid curves) provided the magnetic field dependence at low temperature is $\propto H^2$. The black arrow indicates $T_c$.}
\label{fig3}
\end{figure}

We use a phenomenological model to fit $1/T_1$ in various magnetic fields.  We express the thermal and spatial average over the density of states at the Fermi surface as,
 \begin{equation}
\langle N(0) \rangle = a(H) + b_0 e^{-\Delta_1/k_BT} + c_0 e^{-\Delta_2/k_BT}
\label{eq1}
\end{equation}
where $a(H) = a_0 + a_1H + . . $ and $a_0$ represents  possible contributions from non-magnetic impurities.  The  two gaps,  $\Delta_1$ and   $\Delta_2$, appear in exponential terms with relative weights,  $b_0$ and  $c_0$, as might be expected for the low temperature limit.  Since $1/T_1T \propto \langle N(0)^2 \rangle$, our model for $1/T_1T$ becomes,
\begin{equation}
1/T_1T \propto [ a(H) + b_0 e^{-\Delta_1/k_BT} + c_0 e^{-\Delta_2/k_BT} ]^2.
\label{eq2}
\end{equation}
At low temperatures the two exponential terms are of little  importance and the rate is determined by $a(H)$.  Our numerical analysis provides fits for all of the parameters of the model.  Below $H =16.5$ T, we take them to be magnetic field independent.  However, at this and higher magnetic fields we find that the relative weight of the exponential term from the smaller gap, $b_0$, must be reduced compared to the larger gap weight, $c_0$, in order to fairly represent the data.  As stated previously, these gap parameters are not important in the low temperature limit where we seek to describe the field dependence of the relaxation rate and so we do not ascribe specific importance to this additional field dependence other than it allows us to represent the high temperature behavior in each field.   Nonetheless, we point out that  our results for the temperature dependence at low magnetic field are identical to those from Li {\it et al.}~\cite{li11}  for clean crystals,  except for the lowest temperature point at 4 K.

\begin{figure}[!ht]
%%%%%%%%%%%%%%%%%   F I G U R E  4  %%%%%%%%%%%%%%%%%%
\centerline{\includegraphics[width=0.5\textwidth]{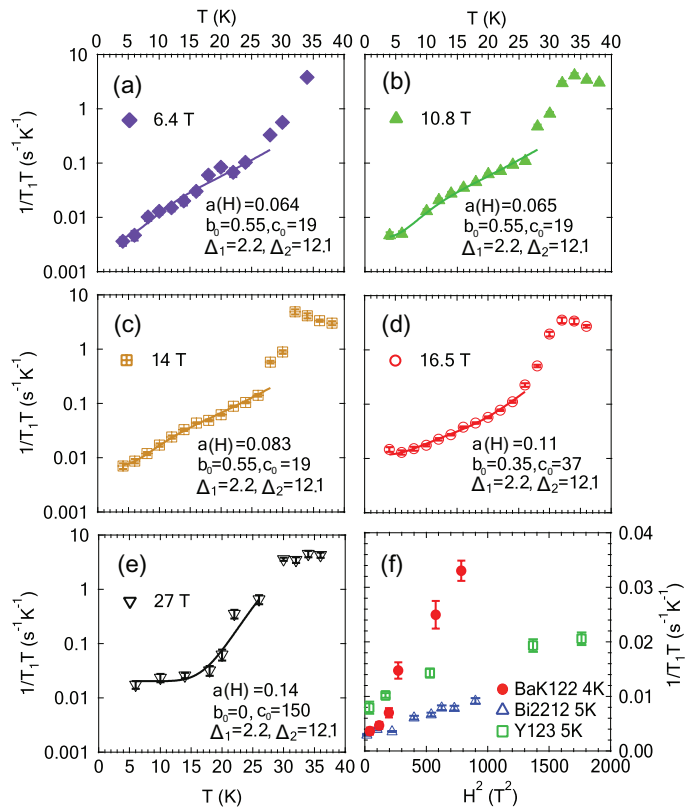}}
%%%%%%%%%%%%%%%%%%%%%%%%%%%%%%%%%%%%%%%%%%%%%
\caption {(a)-(e) $1/^{75}T_1T$ with the best fit curve in each magnetic field. The unit of the $a(H), b_0,$ and $c_0$ is $(s^{-1}K^{-1})$, and the gaps are in meV. (f) $1/^{75}T_1T$ in BaK122 at 4 K,  is shown for comparison with the Zeeman contributions to $1/^{17}T_1T$ in Y123 and Bi2212 at 5 K.  Assuming that the electronic $g$-factor is the same for BaK122 as for the cuprates, we argue that the Zeeman contribution to the field dependence of the average rate we have measured in BaK122 is significantly smaller than from Doppler contributions.  In the case of the cuprates, the Zeeman contributions to the spin-lattice relaxation rate were isolated using frequency resolved measurements performed at the saddle point of the local field distribution (peak of the spectrum) where Doppler contributions cancel based on symmetry of the supercurrents from near-neighbor vortices.
}
\label{fig4}
\end{figure} 

Our analysis in each field is shown in Fig.~\ref{fig4}(a)-(e), where $\Delta_1$ and $\Delta_2$ are 2.1 $\pm$ 0.2 meV and 12.1 $\pm$ 1.4 meV respectively.  The sizes of the gaps correspond well to the sizes of the 3D superconducting gap function from ARPES measurements, 2.07 meV and 12.3 meV.~\cite{xu11} The ratio of the coefficients, $b_0$ and $c_0$, decreases at $H=16.5$ and 27 T, indicating a possible suppression of the smaller superconducting gap, $\Delta_1$, by the external magnetic field.  
The low temperature magnetic field dependence of $1/T_1T$ is given by a(H) shown in  Fig.~\ref{fig4}(f). The  $H^2$ behavior might be associated with Doppler shifted quasiparticles, although the field dependence is different from that predicted by  theory.~\cite{ban11}  It should be noted that the electronic Zeeman interaction also contributes to the quasiparticle energy giving a $H^2$ dependence to $1/T_1T$.~\cite{mit01b}  In Fig.~\ref{fig4}(f) we show the field dependence of $1/T_1T$ for $^{17}$O NMR from YBa$_2$Cu$_3$O$_{7+\delta}$(Y123) aligned powders~\cite{mit01b} and Bi2212 crystals~\cite{mou11} which has been attributed to this Zeeman term.   From comparison with these compounds, allowing for the 27\% larger gyromagnetic ratio of arsenic compared to oxygen, it is reasonable to conclude that the significantly larger field dependence of $1/T_1T$ for BaK122, cannot be attributed to the Zeeman term.  It is notable that the three materials have a  similar value of the spin-lattice relaxation in the limit of zero field.  Although,  to some extent,  this might be fortuitous, it nonetheless suggests that our BaK122 crystal does not have significantly more impurity scattering than these high quality cuprate materials.  Since the NMR linewidth at $T=40$ K is independent of  magnetic field for $H \leq16.5$ T to within 10\%,  we do not associate the field dependence of the rate with magnetic impurities.  However, this possibility can be investigated further by measurement of the frequency-resolved spin-lattice rate which we describe next. 
 
\begin{figure}[!ht]
%%%%%%%%%%%%%%%%%   F I G U R E  5   %%%%%%%%%%%%%%%%%%
\centerline{\includegraphics[width=0.5\textwidth]{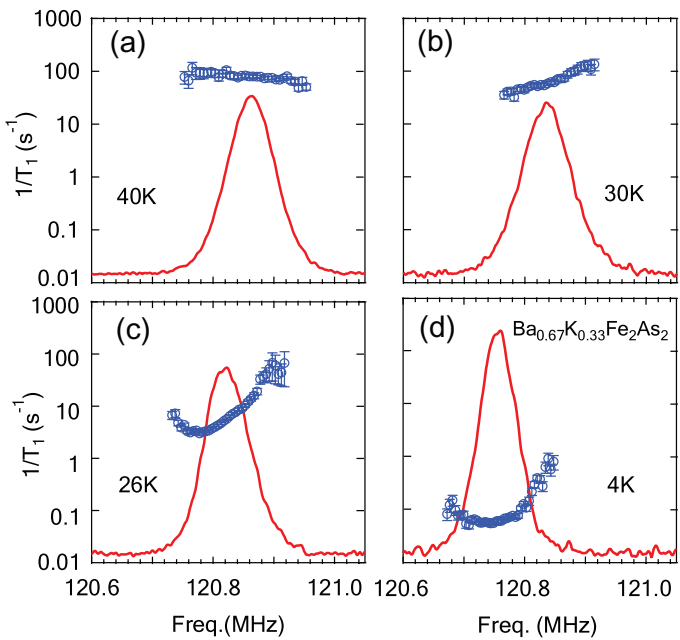}}
%%%%%%%%%%%%%%%%%%%%%%%%%%%%%%%%%%%%%%%%%%%%%
\caption { Spin-lattice relaxation rate across spectrum in the normal state (a) and superconducting states(b),(c),(d) in 16.5 T, with $H||c$-axis. In the normal state, at 40 K, there is no significant frequency dependence in $1/T_1$. However, the rate becomes dependent on frequency as the sample is cooled deep into the superconducting state. }
\label{fig5}
\end{figure}

For unconventional superconductors $1/T_1$ can depend on the position of the probe nucleus relative to the vortex core.~\cite{vol88, tak99, mit01b,mou11} The increase in the supercurrent momentum, ${\bf p}_s$, approaching the core leads to a corresponding increase in the Doppler shift of the energy of quasiparticle excitations, ${\bf v}_F \cdot {\bf p}_s$, where ${\bf v}_F$ is Fermi velocity. The vortex core, having the highest local magnetic field, corresponds to the largest frequency in the NMR spectrum. We have looked for evidence of this spatial dependence of  $1/T_1$ through frequency-resolved, {\it i.e.} spatially resolved, measurements performed across the spectrum, as shown in Fig.~\ref{fig5}.

In the normal state (40 K) we find a flat $1/T_1$ distribution  throughout  the spectrum as expected in the absence of Doppler terms or magnetic impurity contributions to the rate. In the superconducting state, there is an increase of $1/T_1$ with frequency, developing markedly at $T=26$ K with more than an order of magnitude variation across the spectrum.  

\begin{figure}[!ht]
%%%%%%%%%%%%%%%%%   F I G U R E  6   %%%%%%%%%%%%%%%%%%
\centerline{\includegraphics[width=0.5\textwidth]{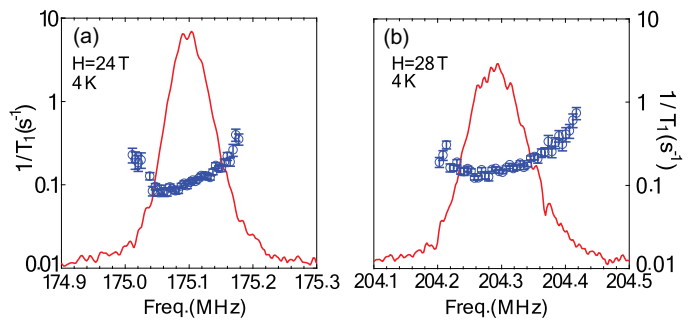}}
%%%%%%%%%%%%%%%%%%%%%%%%%%%%%%%%%%%%%%%%%%%%%
\caption { Spin-lattice relaxation rate across spectrum at 4 K in 24 T (a) and 28 T (b) with $H||c$-axis. The inhomogeneous frequency dependence of the rate is observed in both magnetic fields similar to H = 16.5 T, Fig.\ref{fig5} (d). Additionally, the increase in the high frequency part in the spectra can be understood as an asymmetry from the vortex lattice.}
\label{fig6}
\end{figure}

We note that the linewidth, $\sim 80$ kHz at 4 K in $H=16.5$ T is  somewhat broader than our calculation from Ginzburg-Landau theory using Brandt's algorithm~\cite{bra97} for a perfect vortex lattice, $\sim 23$ kHz.   However, even in a somewhat disordered vortex structure, the high field portion of the spectrum can be associated with nuclei in the vortex core.  This is the case for the distribution in $1/T_1$  observed in Y123, which was attributed to the Doppler shift~\cite{mit01b} of  quasiparticle energy from vortex supercurrents.  Our frequency-resolved measurements of $1/T_1$ in BaK122, Fig.~5, show the existence of a spatially inhomogeneous distribution which onsets with superconductivity.  We ascribe this to the vortex state for which the most likely explanation is a Volovik effect. Another explanation was suggested some years ago to explain observations in superconducting vanadium compounds.~\cite{sil66,sil67}  There it was argued that spin-diffusion from relaxation sources in the vortex core might produce a spatially inhomogeneous distribution of $1/T_1$.  Later measurements and theoretical work by Genack and Redfield~\cite{gen73,gen75} showed that this suggestion was incorrect, and that spin diffusion is quenched on very short time scales owing to depletion of the dipole energy reservoir, an effect even further suppressed with increasing field. We  measured the spin lattice relaxation rates in higher fields,  24 T and 28 T, as shown in Fig.~\ref{fig6}. An inhomogeneous spin-lattice relaxation rate distribution was found similar to that of H = 16.5 T, Fig.~\ref{fig5}, and rules out spin diffusion as a possible mechanism.~\cite{sil66,sil67}

With reports from experiments in cuprates a decade ago~\cite{cur00,mit01b} this mechanism was studied theoretically by Wortis,~\cite{wor98} who came to the same conclusion.  A more detailed discussion has been provided by Mounce {\it et al}.~\cite{mou11b} We point out that in the recent theory~\cite{ban10, ban11} of the Volovik effect in $s\pm$ superconductors the combined effects of the Zeeman interaction and vortex supercurrents have not been taken into account.  Their importance was indicated in the work of Mitrovi\'c {\it et al.} on  YBa$_2$Cu$_3$O$_7$~\cite{mit01b} and might be an important  component missing from the theory. We conclude that our  observations  are most likely a consequence of vortex supercurrents but for which there is not yet a satisfactory theoretical explanation.

In summary, we have studied the $^{75}$As Knight shift and spin-lattice relaxation rate in slightly underdoped Ba$_{0.67}$K$_{0.33}$Fe$_{2}$As$_{2}$ crystals in the superconducting mixed state. We found that  $1/T_1T$ approaches a constant at low temperatures in high magnetic field and is proportional to the square of the field.  Although this is  inconsistent with a theory for the Volovik effect,~\cite{ban10, ban11} our results can be accounted for by a phenomenological model which is based on $s\pm$ symmetry with two isotropic gaps, and non-magnetic impurities. The  distribution of $1/T_1$ across the spectrum resembles that observed in a vortex solid of an unconventional superconductor associated with spatially resolved Doppler contributions to the quasiparticle excitation spectrum. 

We thank Y. Bang, G.E. Volovik, P.J. Hirschfeld, and J.A. Sauls  for helpful discussions.  
Research was supported by the U.S. Department of Energy, Office of Basic Energy Sciences, Division of Materials Sciences and Engineering under Awards DE-FG02-05ER46248 (Northwestern University) and No. DE-FG02-05ER46202 (the University of Tennessee). Work at high magnetic field was performed at the National High Magnetic  Field Laboratory supported by the National Science Foundation and the State of Florida. \\


\begin{thebibliography}{30}
\expandafter\ifx\csname natexlab\endcsname\relax\def\natexlab#1{#1}\fi
\expandafter\ifx\csname bibnamefont\endcsname\relax
  \def\bibnamefont#1{#1}\fi
\expandafter\ifx\csname bibfnamefont\endcsname\relax
  \def\bibfnamefont#1{#1}\fi
\expandafter\ifx\csname citenamefont\endcsname\relax
  \def\citenamefont#1{#1}\fi
\expandafter\ifx\csname url\endcsname\relax
  \def\url#1{\texttt{#1}}\fi
\expandafter\ifx\csname urlprefix\endcsname\relax\def\urlprefix{URL }\fi
\providecommand{\bibinfo}[2]{#2}
\providecommand{\eprint}[2][]{\url{#2}}

\bibitem[{\citenamefont{Volovik}(1988)}]{vol88}
\bibinfo{author}{\bibfnamefont{G.~E.} \bibnamefont{Volovik}},
  \bibinfo{journal}{J. Phys. C.} \textbf{\bibinfo{volume}{21}},
  \bibinfo{pages}{L221} (\bibinfo{year}{1988}).

\bibitem[{\citenamefont{Volovik}(1993)}]{vol93}
\bibinfo{author}{\bibfnamefont{G.~E.} \bibnamefont{Volovik}},
  \bibinfo{journal}{JETP Lett.} \textbf{\bibinfo{volume}{58}},
  \bibinfo{pages}{469} (\bibinfo{year}{1993}).

\bibitem[{\citenamefont{Bang}(2010)}]{ban10}
\bibinfo{author}{\bibfnamefont{Y.}~\bibnamefont{Bang}}, \bibinfo{journal}{Phys.
  Rev. Lett} \textbf{\bibinfo{volume}{104}}, \bibinfo{pages}{217001}
  (\bibinfo{year}{2010}).

\bibitem[{\citenamefont{Bang}(2011)}]{ban11}
\bibinfo{author}{\bibfnamefont{Y.}~\bibnamefont{Bang}} (\bibinfo{year}{2011}),
  \eprint{arXiv.org:1112.0142v2}.

\bibitem[{\citenamefont{Kamihara et~al.}(2008)\citenamefont{Kamihara, Watanabe,
  Hirano, and Hosono}}]{kam08}
\bibinfo{author}{\bibfnamefont{Y.}~\bibnamefont{Kamihara}},
  \bibinfo{author}{\bibfnamefont{T.}~\bibnamefont{Watanabe}},
  \bibinfo{author}{\bibfnamefont{M.}~\bibnamefont{Hirano}}, \bibnamefont{and}
  \bibinfo{author}{\bibfnamefont{H.}~\bibnamefont{Hosono}},
  \bibinfo{journal}{J. Am. Chem. Soc.} \textbf{\bibinfo{volume}{130}},
  \bibinfo{pages}{3296} (\bibinfo{year}{2008}).

\bibitem[{\citenamefont{Nakai et~al.}(2010)\citenamefont{Nakai, Iye, Kitagawa,
  Ishida, Kasahara, Shibauchi, Matsuda, and Terashima}}]{nak10}
\bibinfo{author}{\bibfnamefont{Y.}~\bibnamefont{Nakai}},
  \bibinfo{author}{\bibfnamefont{T.}~\bibnamefont{Iye}},
  \bibinfo{author}{\bibfnamefont{S.}~\bibnamefont{Kitagawa}},
  \bibinfo{author}{\bibfnamefont{K.}~\bibnamefont{Ishida}},
  \bibinfo{author}{\bibfnamefont{S.}~\bibnamefont{Kasahara}},
  \bibinfo{author}{\bibfnamefont{T.}~\bibnamefont{Shibauchi}},
  \bibinfo{author}{\bibfnamefont{Y.}~\bibnamefont{Matsuda}}, \bibnamefont{and}
  \bibinfo{author}{\bibfnamefont{T.}~\bibnamefont{Terashima}},
  \bibinfo{journal}{Phys. Rev. B} \textbf{\bibinfo{volume}{81}},
  \bibinfo{pages}{020503} (\bibinfo{year}{2010}).

\bibitem[{\citenamefont{Ning et~al.}(2010)\citenamefont{Ning, Ahilan, Imai,
  Sefat, McGuire, Sales, Mandrus, Cheng, Shen, and Wen}}]{nin10}
\bibinfo{author}{\bibfnamefont{F.~L.} \bibnamefont{Ning}},
  \bibinfo{author}{\bibfnamefont{K.}~\bibnamefont{Ahilan}},
  \bibinfo{author}{\bibfnamefont{T.}~\bibnamefont{Imai}},
  \bibinfo{author}{\bibfnamefont{A.~S.} \bibnamefont{Sefat}},
  \bibinfo{author}{\bibfnamefont{M.~A.} \bibnamefont{McGuire}},
  \bibinfo{author}{\bibfnamefont{B.~C.} \bibnamefont{Sales}},
  \bibinfo{author}{\bibfnamefont{D.}~\bibnamefont{Mandrus}},
  \bibinfo{author}{\bibfnamefont{P.}~\bibnamefont{Cheng}},
  \bibinfo{author}{\bibfnamefont{B.}~\bibnamefont{Shen}}, \bibnamefont{and}
  \bibinfo{author}{\bibfnamefont{H.-H.} \bibnamefont{Wen}},
  \bibinfo{journal}{Phys. Rev. Lett.} \textbf{\bibinfo{volume}{104}},
  \bibinfo{pages}{037001} (\bibinfo{year}{2010}).

\bibitem[{\citenamefont{Fukazawa et~al.}(2009)\citenamefont{Fukazawa, Yamada,
  Kondo, Saito, Kohori, Kuga, Matsumoto, Nakatsuj, Kito, Shirage
  et~al.}}]{fuk09}
\bibinfo{author}{\bibfnamefont{H.}~\bibnamefont{Fukazawa}},
  \bibinfo{author}{\bibfnamefont{Y.}~\bibnamefont{Yamada}},
  \bibinfo{author}{\bibfnamefont{K.}~\bibnamefont{Kondo}},
  \bibinfo{author}{\bibfnamefont{T.}~\bibnamefont{Saito}},
  \bibinfo{author}{\bibfnamefont{Y.}~\bibnamefont{Kohori}},
  \bibinfo{author}{\bibfnamefont{K.}~\bibnamefont{Kuga}},
  \bibinfo{author}{\bibfnamefont{Y.}~\bibnamefont{Matsumoto}},
  \bibinfo{author}{\bibfnamefont{S.}~\bibnamefont{Nakatsuj}},
  \bibinfo{author}{\bibfnamefont{H.}~\bibnamefont{Kito}},
  \bibinfo{author}{\bibfnamefont{P.~M.} \bibnamefont{Shirage}},
  \bibnamefont{et~al.}, \bibinfo{journal}{J. Phys. Soc. Jpn.}
  \textbf{\bibinfo{volume}{78}}, \bibinfo{pages}{083712}
  (\bibinfo{year}{2009}).

\bibitem[{\citenamefont{Yashima et~al.}(2009)\citenamefont{Yashima, Nishimura,
  Mukuda, Kitaoka, Miyazawa, Shirage, Kihou, Kito, Eisaki, and Iyo}}]{yas09}
\bibinfo{author}{\bibfnamefont{M.}~\bibnamefont{Yashima}},
  \bibinfo{author}{\bibfnamefont{H.}~\bibnamefont{Nishimura}},
  \bibinfo{author}{\bibfnamefont{H.}~\bibnamefont{Mukuda}},
  \bibinfo{author}{\bibfnamefont{Y.}~\bibnamefont{Kitaoka}},
  \bibinfo{author}{\bibfnamefont{K.}~\bibnamefont{Miyazawa}},
  \bibinfo{author}{\bibfnamefont{P.~M.} \bibnamefont{Shirage}},
  \bibinfo{author}{\bibfnamefont{K.}~\bibnamefont{Kihou}},
  \bibinfo{author}{\bibfnamefont{H.}~\bibnamefont{Kito}},
  \bibinfo{author}{\bibfnamefont{H.}~\bibnamefont{Eisaki}}, \bibnamefont{and}
  \bibinfo{author}{\bibfnamefont{A.}~\bibnamefont{Iyo}}, \bibinfo{journal}{J.
  Phys. Soc. Jpn.} \textbf{\bibinfo{volume}{78}}, \bibinfo{pages}{103702}
  (\bibinfo{year}{2009}).

\bibitem[{\citenamefont{Li et~al.}(2011)\citenamefont{Li, Sun, Lin, Su, Hu, and
  Zheng}}]{li11}
\bibinfo{author}{\bibfnamefont{Z.}~\bibnamefont{Li}},
  \bibinfo{author}{\bibfnamefont{D.~L.} \bibnamefont{Sun}},
  \bibinfo{author}{\bibfnamefont{C.~T.} \bibnamefont{Lin}},
  \bibinfo{author}{\bibfnamefont{Y.~H.} \bibnamefont{Su}},
  \bibinfo{author}{\bibfnamefont{J.~P.} \bibnamefont{Hu}}, \bibnamefont{and}
  \bibinfo{author}{\bibfnamefont{G.-q.} \bibnamefont{Zheng}},
  \bibinfo{journal}{Phys. Rev. B} \textbf{\bibinfo{volume}{83}},
  \bibinfo{pages}{140506} (\bibinfo{year}{2011}).

\bibitem[{\citenamefont{Grafe et~al.}(2008)\citenamefont{Grafe, Paar, Lang,
  Curro, Behr, Werner, Hamann-Borrero, Hess, Leps, Klingeler et~al.}}]{gra08}
\bibinfo{author}{\bibfnamefont{H.-J.} \bibnamefont{Grafe}},
  \bibinfo{author}{\bibfnamefont{D.}~\bibnamefont{Paar}},
  \bibinfo{author}{\bibfnamefont{G.}~\bibnamefont{Lang}},
  \bibinfo{author}{\bibfnamefont{N.~J.} \bibnamefont{Curro}},
  \bibinfo{author}{\bibfnamefont{G.}~\bibnamefont{Behr}},
  \bibinfo{author}{\bibfnamefont{J.}~\bibnamefont{Werner}},
  \bibinfo{author}{\bibfnamefont{J.}~\bibnamefont{Hamann-Borrero}},
  \bibinfo{author}{\bibfnamefont{C.}~\bibnamefont{Hess}},
  \bibinfo{author}{\bibfnamefont{N.}~\bibnamefont{Leps}},
  \bibinfo{author}{\bibfnamefont{R.}~\bibnamefont{Klingeler}},
  \bibnamefont{et~al.}, \bibinfo{journal}{Phys. Rev. Lett}
  \textbf{\bibinfo{volume}{101}}, \bibinfo{pages}{047003}
  (\bibinfo{year}{2008}).

\bibitem[{\citenamefont{Yip and Sauls}(1992)}]{yip92}
\bibinfo{author}{\bibfnamefont{S.~K.} \bibnamefont{Yip}} \bibnamefont{and}
  \bibinfo{author}{\bibfnamefont{J.~A.} \bibnamefont{Sauls}},
  \bibinfo{journal}{Phys. Rev. Lett} \textbf{\bibinfo{volume}{69}},
  \bibinfo{pages}{2264} (\bibinfo{year}{1992}).

\bibitem[{\citenamefont{Zhang et~al.}(2011)\citenamefont{Zhang, Wang, Luo,
  Wang, Liu, Zhao, Abernathy, Maier, Marty, Lumsden et~al.}}]{zha11}
\bibinfo{author}{\bibfnamefont{C.}~\bibnamefont{Zhang}},
  \bibinfo{author}{\bibfnamefont{M.}~\bibnamefont{Wang}},
  \bibinfo{author}{\bibfnamefont{H.}~\bibnamefont{Luo}},
  \bibinfo{author}{\bibfnamefont{M.}~\bibnamefont{Wang}},
  \bibinfo{author}{\bibfnamefont{M.}~\bibnamefont{Liu}},
  \bibinfo{author}{\bibfnamefont{J.}~\bibnamefont{Zhao}},
  \bibinfo{author}{\bibfnamefont{D.~L.} \bibnamefont{Abernathy}},
  \bibinfo{author}{\bibfnamefont{T.~A.} \bibnamefont{Maier}},
  \bibinfo{author}{\bibfnamefont{K.}~\bibnamefont{Marty}},
  \bibinfo{author}{\bibfnamefont{M.~D.} \bibnamefont{Lumsden}},
  \bibnamefont{et~al.}, \bibinfo{journal}{Scientific Reports}
  \textbf{\bibinfo{volume}{1}}, \bibinfo{pages}{115} (\bibinfo{year}{2011}).

\bibitem[{\citenamefont{Mitrovi$\acute{c}$
  et~al.}(2001{\natexlab{a}})\citenamefont{Mitrovi$\acute{c}$, Sigmund, and
  Halperin}}]{mit01a}
\bibinfo{author}{\bibfnamefont{V.~F.} \bibnamefont{Mitrovi$\acute{c}$}},
  \bibinfo{author}{\bibfnamefont{E.~E.} \bibnamefont{Sigmund}},
  \bibnamefont{and} \bibinfo{author}{\bibfnamefont{W.~P.}
  \bibnamefont{Halperin}}, \bibinfo{journal}{Phys. Rev. B.}
  \textbf{\bibinfo{volume}{64}}, \bibinfo{pages}{024520}
  (\bibinfo{year}{2001}{\natexlab{a}}).

\bibitem[{\citenamefont{Chen et~al.}(2007)\citenamefont{Chen, Halperin,
  Guptasarma, Hinks, Mitrovic, Reyes, and Kuhns}}]{che07}
\bibinfo{author}{\bibfnamefont{B.}~\bibnamefont{Chen}},
  \bibinfo{author}{\bibfnamefont{W.~P.} \bibnamefont{Halperin}},
  \bibinfo{author}{\bibfnamefont{P.}~\bibnamefont{Guptasarma}},
  \bibinfo{author}{\bibfnamefont{D.~G.} \bibnamefont{Hinks}},
  \bibinfo{author}{\bibfnamefont{V.~F.} \bibnamefont{Mitrovic}},
  \bibinfo{author}{\bibfnamefont{A.~P.} \bibnamefont{Reyes}}, \bibnamefont{and}
  \bibinfo{author}{\bibfnamefont{P.~L.} \bibnamefont{Kuhns}},
  \bibinfo{journal}{Nature Physics} \textbf{\bibinfo{volume}{3}},
  \bibinfo{pages}{239} (\bibinfo{year}{2007}).

\bibitem[{\citenamefont{Oh et~al.}(2011)\citenamefont{Oh, Mounce, Mukhopadhyay,
  Halperin, Vorontsov, Bud'ko, Canfield, Furukawa, Reyes, and Kuhns}}]{oh11}
\bibinfo{author}{\bibfnamefont{S.}~\bibnamefont{Oh}},
  \bibinfo{author}{\bibfnamefont{A.~M.} \bibnamefont{Mounce}},
  \bibinfo{author}{\bibfnamefont{S.}~\bibnamefont{Mukhopadhyay}},
  \bibinfo{author}{\bibfnamefont{W.~P.} \bibnamefont{Halperin}},
  \bibinfo{author}{\bibfnamefont{A.~B.} \bibnamefont{Vorontsov}},
  \bibinfo{author}{\bibfnamefont{S.~L.} \bibnamefont{Bud'ko}},
  \bibinfo{author}{\bibfnamefont{P.~C.} \bibnamefont{Canfield}},
  \bibinfo{author}{\bibfnamefont{Y.}~\bibnamefont{Furukawa}},
  \bibinfo{author}{\bibfnamefont{A.~P.} \bibnamefont{Reyes}}, \bibnamefont{and}
  \bibinfo{author}{\bibfnamefont{P.~L.} \bibnamefont{Kuhns}},
  \bibinfo{journal}{Phys. Rev. B.} \textbf{\bibinfo{volume}{83}},
  \bibinfo{pages}{214501} (\bibinfo{year}{2011}).

\bibitem[{\citenamefont{Mukhopadhyay et~al.}(2009)\citenamefont{Mukhopadhyay,
  Oh, Mounce, Lee, Halperin, Ni, Bud'ko, Canfield, Reyes, and Kuhns}}]{muk09}
\bibinfo{author}{\bibfnamefont{S.}~\bibnamefont{Mukhopadhyay}},
  \bibinfo{author}{\bibfnamefont{S.}~\bibnamefont{Oh}},
  \bibinfo{author}{\bibfnamefont{A.~M.} \bibnamefont{Mounce}},
  \bibinfo{author}{\bibfnamefont{M.}~\bibnamefont{Lee}},
  \bibinfo{author}{\bibfnamefont{W.~P.} \bibnamefont{Halperin}},
  \bibinfo{author}{\bibfnamefont{N.}~\bibnamefont{Ni}},
  \bibinfo{author}{\bibfnamefont{S.~L.} \bibnamefont{Bud'ko}},
  \bibinfo{author}{\bibfnamefont{P.~C.} \bibnamefont{Canfield}},
  \bibinfo{author}{\bibfnamefont{A.~P.} \bibnamefont{Reyes}}, \bibnamefont{and}
  \bibinfo{author}{\bibfnamefont{P.~L.} \bibnamefont{Kuhns}},
  \bibinfo{journal}{New J. Phys.} \textbf{\bibinfo{volume}{11}},
  \bibinfo{pages}{055002} (\bibinfo{year}{2009}).

\bibitem[{\citenamefont{Xu et~al.}(2011)\citenamefont{Xu, Huang, Cui, Razzoli,
  Radovic, Shi, Chen, Zheng, Wang, Zhang et~al.}}]{xu11}
\bibinfo{author}{\bibfnamefont{Y.-M.} \bibnamefont{Xu}},
  \bibinfo{author}{\bibfnamefont{Y.-B.} \bibnamefont{Huang}},
  \bibinfo{author}{\bibfnamefont{X.-Y.} \bibnamefont{Cui}},
  \bibinfo{author}{\bibfnamefont{E.}~\bibnamefont{Razzoli}},
  \bibinfo{author}{\bibfnamefont{M.}~\bibnamefont{Radovic}},
  \bibinfo{author}{\bibfnamefont{M.}~\bibnamefont{Shi}},
  \bibinfo{author}{\bibfnamefont{G.-F.} \bibnamefont{Chen}},
  \bibinfo{author}{\bibfnamefont{P.}~\bibnamefont{Zheng}},
  \bibinfo{author}{\bibfnamefont{N.-L.} \bibnamefont{Wang}},
  \bibinfo{author}{\bibfnamefont{C.-L.} \bibnamefont{Zhang}},
  \bibnamefont{et~al.}, \bibinfo{journal}{Nature Physics}
  \textbf{\bibinfo{volume}{7}}, \bibinfo{pages}{198} (\bibinfo{year}{2011}).

\bibitem[{\citenamefont{Mitrovi$\acute{c}$
  et~al.}(2001{\natexlab{b}})\citenamefont{Mitrovi$\acute{c}$, Sigmund,
  Eschrig, Bachman, Halperin, Reyes, Kuhns, and Moulton}}]{mit01b}
\bibinfo{author}{\bibfnamefont{V.~F.} \bibnamefont{Mitrovi$\acute{c}$}},
  \bibinfo{author}{\bibfnamefont{E.~E.} \bibnamefont{Sigmund}},
  \bibinfo{author}{\bibfnamefont{E.}~\bibnamefont{Eschrig}},
  \bibinfo{author}{\bibfnamefont{H.~N.} \bibnamefont{Bachman}},
  \bibinfo{author}{\bibfnamefont{W.~P.} \bibnamefont{Halperin}},
  \bibinfo{author}{\bibfnamefont{A.~P.} \bibnamefont{Reyes}},
  \bibinfo{author}{\bibfnamefont{P.}~\bibnamefont{Kuhns}}, \bibnamefont{and}
  \bibinfo{author}{\bibfnamefont{W.~G.} \bibnamefont{Moulton}},
  \bibinfo{journal}{Nature} \textbf{\bibinfo{volume}{413}},
  \bibinfo{pages}{501} (\bibinfo{year}{2001}{\natexlab{b}}).

\bibitem[{\citenamefont{Mounce et~al.}(2011{\natexlab{a}})\citenamefont{Mounce,
  Oh, Mukhopadhyay, Halperin, Reyes, Kuhns, Fujita, Ishikado, and
  Uchida}}]{mou11}
\bibinfo{author}{\bibfnamefont{A.~M.} \bibnamefont{Mounce}},
  \bibinfo{author}{\bibfnamefont{S.}~\bibnamefont{Oh}},
  \bibinfo{author}{\bibfnamefont{S.}~\bibnamefont{Mukhopadhyay}},
  \bibinfo{author}{\bibfnamefont{W.~P.} \bibnamefont{Halperin}},
  \bibinfo{author}{\bibfnamefont{A.~P.} \bibnamefont{Reyes}},
  \bibinfo{author}{\bibfnamefont{P.~L.} \bibnamefont{Kuhns}},
  \bibinfo{author}{\bibfnamefont{K.}~\bibnamefont{Fujita}},
  \bibinfo{author}{\bibfnamefont{M.}~\bibnamefont{Ishikado}}, \bibnamefont{and}
  \bibinfo{author}{\bibfnamefont{S.}~\bibnamefont{Uchida}},
  \bibinfo{journal}{Phys. Rev. Lett} \textbf{\bibinfo{volume}{106}},
  \bibinfo{pages}{057003} (\bibinfo{year}{2011}{\natexlab{a}}).

\bibitem[{\citenamefont{Takigawa et~al.}(1999)\citenamefont{Takigawa, Ichioka,
  and Machida}}]{tak99}
\bibinfo{author}{\bibfnamefont{M.}~\bibnamefont{Takigawa}},
  \bibinfo{author}{\bibfnamefont{M.}~\bibnamefont{Ichioka}}, \bibnamefont{and}
  \bibinfo{author}{\bibfnamefont{K.}~\bibnamefont{Machida}},
  \bibinfo{journal}{Phys. Rev. Lett.} \textbf{\bibinfo{volume}{83}},
  \bibinfo{pages}{3057} (\bibinfo{year}{1999}).

\bibitem[{\citenamefont{Brandt}(1997)}]{bra97}
\bibinfo{author}{\bibfnamefont{E.~H.} \bibnamefont{Brandt}},
  \bibinfo{journal}{Phys. Rev. Lett.} \textbf{\bibinfo{volume}{78}},
  \bibinfo{pages}{2208} (\bibinfo{year}{1997}).

\bibitem[{\citenamefont{Silbernagel et~al.}(1966)\citenamefont{Silbernagel,
  Weger, and Wernick}}]{sil66}
\bibinfo{author}{\bibfnamefont{B.}~\bibnamefont{Silbernagel}},
  \bibinfo{author}{\bibfnamefont{M.}~\bibnamefont{Weger}}, \bibnamefont{and}
  \bibinfo{author}{\bibfnamefont{J.}~\bibnamefont{Wernick}},
  \bibinfo{journal}{Phys. Rev. Lett.} \textbf{\bibinfo{volume}{17}},
  \bibinfo{pages}{384} (\bibinfo{year}{1966}).

\bibitem[{\citenamefont{Silbernagel et~al.}(1967)\citenamefont{Silbernagel,
  Weger, Clark, and Wernick}}]{sil67}
\bibinfo{author}{\bibfnamefont{B.~G.} \bibnamefont{Silbernagel}},
  \bibinfo{author}{\bibfnamefont{M.}~\bibnamefont{Weger}},
  \bibinfo{author}{\bibfnamefont{W.~G.} \bibnamefont{Clark}}, \bibnamefont{and}
  \bibinfo{author}{\bibfnamefont{J.~H.} \bibnamefont{Wernick}},
  \bibinfo{journal}{Phys. Rev.} \textbf{\bibinfo{volume}{153}},
  \bibinfo{pages}{535} (\bibinfo{year}{1967}).

\bibitem[{\citenamefont{Genack and Redfield}(1973)}]{gen73}
\bibinfo{author}{\bibfnamefont{A.}~\bibnamefont{Genack}} \bibnamefont{and}
  \bibinfo{author}{\bibfnamefont{A.}~\bibnamefont{Redfield}},
  \bibinfo{journal}{Phys. Rev. Lett.} \textbf{\bibinfo{volume}{31}},
  \bibinfo{pages}{1204} (\bibinfo{year}{1973}).

\bibitem[{\citenamefont{Genack and Redfield}(1975)}]{gen75}
\bibinfo{author}{\bibfnamefont{A.}~\bibnamefont{Genack}} \bibnamefont{and}
  \bibinfo{author}{\bibfnamefont{A.}~\bibnamefont{Redfield}},
  \bibinfo{journal}{Phys. Rev. B} \textbf{\bibinfo{volume}{12}},
  \bibinfo{pages}{78} (\bibinfo{year}{1975}).

\bibitem[{\citenamefont{Curro et~al.}(2000)\citenamefont{Curro, Milling, Haase,
  and Slichter}}]{cur00}
\bibinfo{author}{\bibfnamefont{N.~J.} \bibnamefont{Curro}},
  \bibinfo{author}{\bibfnamefont{C.}~\bibnamefont{Milling}},
  \bibinfo{author}{\bibfnamefont{J.}~\bibnamefont{Haase}}, \bibnamefont{and}
  \bibinfo{author}{\bibfnamefont{C.~P.} \bibnamefont{Slichter}},
  \bibinfo{journal}{Phys. Rev. B} \textbf{\bibinfo{volume}{62}},
  \bibinfo{pages}{3473} (\bibinfo{year}{2000}).

\bibitem[{\citenamefont{Wortis}(1998)}]{wor98}
\bibinfo{author}{\bibfnamefont{R.}~\bibnamefont{Wortis}}, Ph.D. thesis,
  \bibinfo{school}{University of Illinois Champaign Urbana}
  (\bibinfo{year}{1998}); \bibinfo{author}{\bibfnamefont{R.}~\bibnamefont{Wortis}},
  \bibinfo{author}{\bibfnamefont{A.~J.} \bibnamefont{Berlinsky}},
  \bibnamefont{and} \bibinfo{author}{\bibfnamefont{C.}~\bibnamefont{Kallin}},
  \bibinfo{journal}{Phys. Rev. B} \textbf{\bibinfo{volume}{61}},
  \bibinfo{pages}{12342} (\bibinfo{year}{2000}).

\bibitem[{\citenamefont{Mounce et~al.}(2011{\natexlab{b}})\citenamefont{Mounce,
  Oh, and Halperin}}]{mou11b}
\bibinfo{author}{\bibfnamefont{A.}~\bibnamefont{Mounce}},
  \bibinfo{author}{\bibfnamefont{S.}~\bibnamefont{Oh}}, \bibnamefont{and}
  \bibinfo{author}{\bibfnamefont{W.}~\bibnamefont{Halperin}},
  \bibinfo{journal}{Front. Phys.} \textbf{\bibinfo{volume}{6}},
  \bibinfo{pages}{450} (\bibinfo{year}{2011}{\natexlab{b}}).

\end{thebibliography}
\end{document}